\documentclass[aps,prl,twocolumn]{revtex4}

\usepackage{amsfonts}%
\usepackage{amsmath}%
\usepackage{amssymb}%
\usepackage{graphicx}

\begin{document}

\title{Any nonlocal model assuming ``local parts'' conflicts with relativity}
\author{Antoine Suarez}
\affiliation{Center for Quantum Philosophy, Berninastrasse 85, 8057 Zurich,Switzerland/suarez@leman.ch}
\pacs{PACS number}
\date{April 16, 2013}
\begin{abstract}
It is argued that any nonlocal model producing ``local parts'' can be reproduced by ``multisimultaneity'' and therefore (because of \cite{Scarani13}) conflicts not only with quantum mechanics but also with relativity. This result means in particular that the very structure of space-time (relativity) requires influences coming from outside space-time.
\end{abstract}
\maketitle

\ \\
\textbf{Introduction}.\textemdash The violation of Bell inequalities proves that the quantum predictions in 2-particle entanglement experiments cannot be explained by influences propagating at a velocity bounded by the speed of light~\cite{nouvellecuisine}. After the experimental demonstration of such a violation there have been several proposals to explain quantum nonlocality by mechanisms that lead to local parts, i.e.: disappearance of the correlations under certain testable conditions. The first consistently developed models of this kind have been ``finite speed'' (proposed by Philippe Eberhard \cite{Eberhardt89}) and ``multisimultaneity'' (proposed by Valerio Scarani and myself \cite{asvs97, S97}). Both have been tested and falsified by experiment \cite{ZBGT01,SZGS02,SZGS03,Salart08}, and recently proved to conflict with relativity \cite{Scarani13, Barnea13}.

After finite speed and multisimultaneity two other nonlocal models assuming local parts have been proposed and received special attention: The Leggett's model \cite{le, gro} and the Colbeck-Renner (C\&R) one \cite{core, Colbeck11, Stuart12}.

In this paper I present a new version of the argument in \cite{Scarani13} proving that multisimultaneity conflicts with relativity. Thereafter I show that finite-speed, Leggett and C\&R models reduce to multisimultaneity, and conclude that any nonlocal model assuming local parts conflict with relativity.

For analysis it is useful to distinguish between theories assuming decision at detection (like standard quantum mechanics) and those assuming decision at the beam-splitters (like Bohmian mechanics). \cite{Suarez12}

\begin{figure}
\includegraphics[width=0.7\columnwidth]{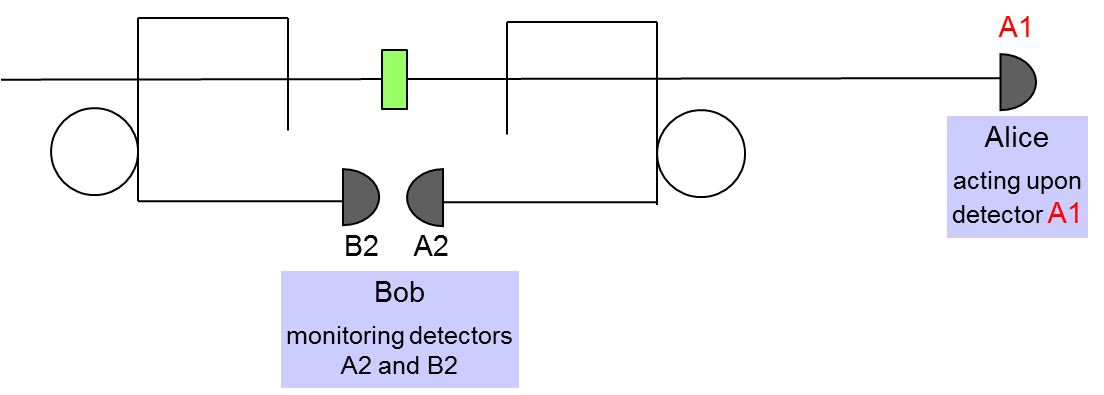}
\caption{\textbf{Experiment to demonstrate conflict with relativity assuming decision at detection:} A1 in red means that Alice can break the nonlocal coordination between detector A1 and detector B2 by changing the settings of A1. Then the rate of the joint outcomes (A2, B2) depends on the settings of detector A1, and Alice (operating upon A1) can message to Bob (watching at B2 and A2) faster-than-light.}
\label{f1}
\end{figure}

\ \\
\textbf{Assuming decision at detection}.\textemdash If one follows \emph{standard} quantum mechanics and assumes decision at detection \cite{Suarez12, Guerreiro12}, then one can prove (already in single-particle experiments) that any disappearance of nonlocal correlations implies straightforwardly violation of the conservation of energy in each single quantum event \cite{Suarez12}, and therefore nonlocal models with local parts can be considered ruled out by the experiment presented in~\cite{Guerreiro12}.

Additionally, by means of the 2-particle experiment sketched in Figure \ref{f1} (which generalizes an argument in \cite{ZBGT01}) one can show that any arrangement of the detectors thwarting the nonlocal coordination between Alice's detectors and Bob's ones would allow Alice phoning to Bob faster-than-light.

\ \\
\textbf{Assuming decision at the beam-splitters}.\textemdash This amounts to work with models that in principle accept de Broglie's \emph{``empty waves''} (i.e.: entities that propagate in the space-time but notoriously cannot be directly accessed). In this context the ``measurement settings''  refer usually to the orientation of the polarizing beam-splitters or the optical path difference of interferometers, depending on the experiment.

\begin{figure}
\includegraphics[width=0.9\columnwidth]{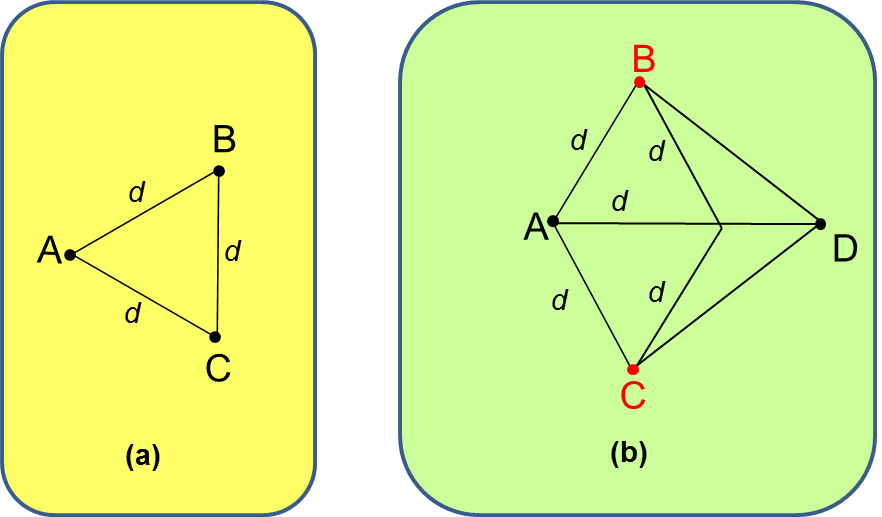}
\caption{\textbf{Experiment to demonstrate conflict with relativity assuming decision at the beam-splitters:} A (Alice), B (Bob), and C (Charlie) represent the 3 beam-splitters measuring the tripartite quantum state described in \cite{Barnea13}. With configuration (\textbf{a}) the model is supposed to reproduce the quantum mechanical predictions. With configuration (\textbf{b}) the nonlocal coordination between B and C breaks down and the model deviates from quantum mechanics: The marginal Bob-Charlie depends on Alice's settings. The distance $d$ refers to models and alignments imposing a critical distance $L$ at which the nonlocal coordination breaks down ($d<L<BC$). Since $AD>BD=CD$ it is always possible to find a point D which lies in the future lightcone of each of Bob's and Charlie's measurements (at B and C), but lies outside the future lightcone of the Alice's measurement (at A).}
\label{f2}
\end{figure}

Suppose the 3-particle entanglement experiment sketched in Figure \ref{f2}. The configuration (a) represents the case where the alternative model reproduces the quantum mechanical predictions. The configuration (b) represents the case where according to the model the nonlocal coordination between the beam-splitters B and C breaks down (symbolized in Figure \ref{f2}b by labeling the corresponding beam-splitters in red). For all marginals which do not contain two beam-splitters in red, the model is supposed to reproduce the quantum correlations.

On the basis of a convenient 3-particle quantum state it has been proved that the nonlocal models assuming finite speed (Eberhard models) lead to faster-than-light communication \cite{Barnea13}. This proof is the tripartite version of the achievement for the four-party scenario \cite{Bancal12}.

The proofs in \cite{Barnea13} and \cite{Bancal12} rely on an intermediate result that is actually independent of the assumption of ``finite speed''. In the tripartite case this result reads: \emph{The assumption of local parts necessarily imposes that the marginal of the joint outcomes Bob-Charlie depends on the settings of Alice} (Figure \ref{f2}b)).

The argument in \cite{Scarani13} can now easily be simplified and implemented in the tripartite case as depicted in Figure \ref{f2}b): Since the $AD>DB=DC$ it is always possible to choose a point D which is in the future lightcone of the measurements at each of the beam splitters B and C, but lies outside the future lightcone of the measurement at A. Then Alice can \emph{nonlocally} code a message by modifying the marginal BC; Bob and Charlie can communicate \emph{locally} their outcomes to an observer in D, who thereby will receive Alice's message still faster-than-light. Notice that neither Bob nor Charlie, are capable of receiving Alice's message faster-than-light.

In summary, the possibility of switching off the nonlocal coordination between two beam-splitters (assumption of ``local parts'') conflicts with relativity; the models finite speed and multisimultaneity can be considered particular cases of the assumption of ``local parts''. This result for models assuming ``empty waves'' (decision at the beam-splitter) is the counterpart of the demonstration in Figure \ref{f1} for models assuming decision at detection.

\ \\
\textbf{Falsification of multisimultaneity implies falsification of finite-speed}.\textemdash Actually any before-before experiment includes a test of the finite-speed assumption.

Suppose a model assuming a finite speed $v>c$ in some universal preferred frame. Then the lower bound $L$ for the distance to ensure disappearance of the nonlocal correaltions is given by:

\begin{equation}
L>v \left| \Delta t\right|
\label{L}
\end{equation}

\ \\
where $\Delta t$ is the time difference between the arrivals of each particle pair at the corresponding beam-splitters, and measures the precision of alignment.

Suppose now a before-before experiment with two beam-splitters moving away from each other at velocity $v_{bb}$. The criterion for disappearance of the nonlocal coordination at a distance $L$ reads \cite{asvs97}:
\begin{equation}
\left| \Delta t\right| <\frac{v_{bb}}{c^{2}}L
\label{dt}
\end{equation}

From (\ref{L}) and (\ref{dt}) it follows:

\begin{equation}
v_{bb}=\frac{c^{2}}{L} \left| \Delta t\right|=\frac{c^{2}}{v}
\label{vbb}
\end{equation}

Suppose now that a before-before experiment with $v_{bb}$ as in (\ref{vbb}) and separation $L$ as in (\ref{L}) demonstrates nonlocal correlations. Suppose now that one stops the movement of the beam-splitters. One should coherently conclude that the nonlocal correlations remain, and thereby the experiment falsifies also the assumption of hidden influences propagating at speed $v$ as in (\ref{vbb}). So for instance, the before-before experiment performed in \cite{SZGS02, SZGS03} rules out the assumption that the hidden nonlocal influences propagate at a velocity $v\simeq10^5 c$ in the laboratory frame.

The interesting point is now the following: If one accepts that the before-before experiment in \cite{SZGS02, SZGS03} falsifies multisimultaneity for any possible value of $v_{bb}$, one should coherently assume as well, that the same experiment falsifies the finite-speed model for any possible speed $v$. Notice by contrast that the experiment in \cite{Salart08} establishes only a lower bound for the finite-speed.

\ \\
\textbf{Leggett and C\&R reduce to multisimultaneity}.\textemdash All these three models share in the assumption that there is some mechanism or operation allowing us to switch off the nonlocal coordination, and thereby produce a testable deviation from quantum mechanics.

In multisimultaneity the operation by which the experimenter can switch off the nonlocal coordination is well defined and easy to perform: It consist in setting the choice devices (detectors or beam-splitters) at distances and velocities leading to a before-before timing. By contrast Leggett and C\&R models state only that such a mechanism exist but do not tell us how to put it to work. This leads to a remarkable difference about the experimental protocol to test the models:

To test multisimultaneity the experimenter arranges two different configurations (like in Figure \ref{f2}): One in which the model makes the same prediction as quantum mechanics (before-after timing), and another configuration in which the model predicts disappearance of the nonlocal coordination (before-before timing). We denote S1 the set of joint-outcomes measured in the first configuration, and S2 the set measured in the second one, where S2 contains exclusively joint-outcomes that are locally correlated (for instance because of ``shared randomness''). Then the experimenter confirms that the set S1 violates the CHSH inequality, and test whether this is or not the case for S2.

To test Leggett and C\&R models the experimenter produces only one unique set of outcomes S, consisting of both nonlocal and local parts. Because of the nonlocal content the set S violates the CHSH inequality, but because of the local content it fulfills some other inequality which is violated by standard quantum mechanics (Leggett's inequalities or an N-chained Bell inequality beyond certain N in case of C\&R). Meanwhile it has been showed that Leggett's model can also be described in terms of fulfillment of chained Bell inequalities \cite{Colbeck11, Stuart12, Navascues}. Thus the trial for this kind of models consists in assessing whether the set S fulfills a N-chained Bell inequality beyond a certain N ($2<N<\infty$).

Suppose now that in case of Leggett and C\&R models the experimenter gets hold of the (according to the model) existing accessible mechanism to switch off the nonlocal coordination, like in the case of multisimultaneity. Would he/she still test these models following the protocol described in the preceding paragraph? Obviously not. He/she would proceed like in the case of the experiments testing multisimultaneity, that is, producing a set S2 containing only presumed local parts and testing wether it violates the CHSH inequality or not.

Conversely, in case of multisimultaneity one could easily produce a set containing both local parts (with weight \emph{p}) and nonlocal parts (with weight $1-p$) simply by switching off and on the mechanism producing the before-before timing. The mixture produced this way will fulfill a N-chained Bell inequality beyond a certain $N$ depending on the weight $p$ \cite{Pironio06, core}, whereas quantum mechanics predicts a violation for $N$ arbitrarily large. However nobody would come to the idea of testing multisimultaneity by producing such a set and assessing a possible deviation from quantum mechanics by means of a chained Bell inequality. And if by whatever reason he/she does it, he/she would not prove anything about nonlocality different from what he would prove by verifying violation of the CHSH inequality.

This means that Leggett and C\&R assume as a premise the theorem they claim to proof (i.e. deviation from quantum mechanics). Therefore the addressed models are essentially of the same type as multisimultaneity, and  Leggett-type experiments assuming non-covariant (time-ordered) nonlocal hidden influences (``nonlocal realism'') \cite{gro} do not prove anything about alternative nonlocal models different from what the before-before experiment with \emph{moving beam-splitters} did prove \cite{as08}.

Consider now covariant Legett or C\&R models, which by axiom (called of ``free choice'') exclude non-covariant (time-ordered) hidden influences like those invoked by Bohmian mechanics and multisimultaneity \cite{core, Colbeck11, Stuart12, cb}. Thereby the models exclude decision at the beam-splitter, and accept decision at detection. But then the models imply violation of the conservation of energy \cite{Suarez12}, and are ruled out by the experiment in \cite{Guerreiro12}. Furthermore they are proved to conflict with relativity by the argument represented in Figure \ref{f1}, and ruled out by the experiment presented in \cite{ZBGT01}. This means that the arguments and results presented in \cite{core, Stuart12, cb} may lead to a useful cryptographic tool but do not prove anything about nonlocal models with local parts beyond what the before-before experiment with \emph{moving detectors} did prove \cite{ZBGT01}.

In summary, nonlocal models with local parts deviate from quantum mechanics by definition, not by theorem. Arguing that such models deviate from quantum mechanics begs the question. The interesting things to do with them are: 1) To test them  vs. quantum mechanics by experiment. 2) To prove (by reasoning) that they conflict with relativity (in the sense that they imply the possibility of phoning faster than light). The latter is not done in \cite{Colbeck11} and \cite{Stuart12}. But since the models can be considered equivalent to multisimultaneity in terms of operational predictions, they are proved to conflict with relativity by the arguments presented in this paper.

In this context it is also worth to remark that mutisimultaneity can be described as a model postulating a mechanism that allow us to produce two different statistical distributions for the same quantum state: the nonlocal quantum mechanical distribution, and a second local one. From this point of view multisimultaneity can be considered a representative of the so (equivocally) called ``statistical interpretation'' or PBR models \cite{pbr, rc12}. But one could consider a different class of (PBR-like) models associating to the quantum state different nonlocal distributions, i.e. without local parts. Claiming to prove that such models deviate from quantum mechanics means again begging the question. By contrast it may be interesting to ask whether such models conflict with relativity. It is obvious that under assumption of ``decision at detection'' the argument represented in Figure \ref{f1} proves also that any PBR-like model conflicts with relativity. However to prove the same under ``assumption of decision at the beam-splitter'' remains an open question.

\ \\
\textbf{Conclusion}.\textemdash  I would like to conclude with three remarks:

1) My motivation in proposing \emph{multisimultaneity} was describing quantum nonlocality by means of hidden influences acting somewhat within space-time, keeping as far as possible to the ``prejudice'' of time-ordered causality. As we see, this ``prejudice'' leads to faster-than-light communication and thereby is at odds with relativity, that is the very structure of space-time. Ironically, it seems that to have a world with space-time we have to accept coordination coming from outside space-time.

2) The de Broglie's ``local empty wave'' permits to escape the standard nonlocality at detection in single-particle experiments \cite{Suarez12}. The fact that in 2-particle experiments nonlocality reappears through the violation of Bell's inequalities is undoubtedly astonishing. But even more astonishing is this other fact: While models with ``local parts'' can be proved to conflict with relativity quite easily within the standard (collapse at detection) quantum mechanics (as shown in Figure \ref{f1}), the proof is pretty demanding within the de Broglie-Bohmian (``empty wave'') mechanics \cite{Bancal12, Scarani13, Barnea13}. May be this a sign that nonlocality at detection is more basic than Bell's nonlocality? In any case is a sign that it is worth to clarify the relationship between this two kinds of nonlocalities, and probably necessary in order to found quantum mechanics consistently on axioms \cite{Suarez10}.

3) It has been argued that with ``empty waves'' one cannot consistently escape the ``parallell lives" version of ``many worlds'' \cite{Suarez13}. If one accepts this conclusion, then ``decision at the beam splitters'' does not allow us to incorporate consistently nonlocality after all, and the whole work of proving that nature is nonlocal makes sense only in the context of ``decision at detection''. But then the obvious way to do it is by means of the experiment presented in \cite{Guerreiro12}, which furthermore is apparently loophole free \cite{Suarez12a}.

\ \\
\emph{Acknowledgments}

I am gratefull to Antonio Acin, Nicolas Gisin, Stefano Pironio and Valerio Scarani for stimulating comments, and to he Social Trends Institute (Barcelona and New York) for support.

\end{document}